\documentclass[acmsmall, nonacm=true]{acmart}

\usepackage{import}

\usepackage{xspace}

\usepackage{xcolor}

\definecolor{bkgd}{RGB}{240,242,246}
\definecolor{eclipseBlue}{RGB}{42,0.0,255}
\definecolor{eclipseGreen}{RGB}{63,127,95}
\definecolor{eclipsePurple}{RGB}{127,0,85}
\definecolor{eclipseRed}{RGB}{223,26,65}
\definecolor{sublimegreen}{RGB}{112,200,0}
\definecolor{purple}{RGB}{255, 0, 255}

\definecolor{8ECFC9}{RGB}{142,207,201}
\definecolor{FFBE7A}{RGB}{255,190,122}
\definecolor{FA7F6F}{RGB}{250,127,111}
\definecolor{82BOD2}{RGB}{130,176,210}
\definecolor{0067D2}{RGB}{0,103,210}
\definecolor{codegreen}{rgb}{0,0.6,0}
\definecolor{yellow}{RGB}{195, 113, 0}

\usepackage{amssymb}
\usepackage{amsmath}
\usepackage{multirow}
\usepackage{multicol}
\usepackage{subfigure}

\usepackage{realboxes}

\usepackage{enumitem}

\usepackage{algpseudocode}

\usepackage{braket}

\usepackage{url}

\usepackage{natbib}

\usepackage{lipsum}

\usepackage{bm}

\usepackage{enumitem}

\usepackage{graphicx}
\usepackage{pythonhighlight}
\usepackage{listings}

\usepackage{tikz}
\usetikzlibrary{decorations.pathreplacing,decorations.pathmorphing}

\usepackage{siunitx}

\usepackage{parcolumns}

\usepackage{float}
\newfloat{lstfloat}{htbp}{lop}
\floatname{lstfloat}{Listing}

\RequirePackage{listings}
\RequirePackage{xcolor}

\lstdefinelanguage{Python}
{
    showtabs=true,
    tab=,
    tabsize=2,
    basicstyle=\linespread{1.0}\small\ttfamily,
    stringstyle=\color{stringcolour},
    showstringspaces=false,
    alsoletter={1234567890},
    otherkeywords={\%, \}, \{}, 
    keywordstyle=\color{keywordcolour}\bfseries,
    emph={and,break,class,continue,def,yield,del,elif ,else,%
    except,exec,finally,for,from,global,if,import,in,%
    lambda,not,or,pass,print,raise,return,try,while,assert,with},
    emphstyle=\color{blue}\bfseries,
    emph={[2]True, False, None},
    emphstyle=[2]\color{keywordcolour},
    emph={[3]object,type,isinstance,copy,deepcopy,zip,enumerate,reversed,list,set,len,dict,tuple,xrange,append,execfile,real,imag,reduce,str,repr},
    emphstyle=[3]\color{commandcolour},
    emph={Exception,NameError,IndexError,SyntaxError,TypeError,ValueError,OverflowError,ZeroDivisionError},
    emphstyle=\color{exceptioncolour}\bfseries,
    morecomment=[l]{\#}, 
    commentstyle=\color{commentcolour}\slshape,
    emph={[4]ode, fsolve, sqrt, exp, sin, cos,arctan, arctan2, arccos, pi,  array, norm, solve, dot, arange, isscalar, max, sum, flatten, shape, reshape, find, any, all, abs, plot, linspace, legend, quad, polyval,polyfit, hstack, concatenate,vstack,column_stack,empty,zeros,ones,rand,vander,grid,pcolor,eig,eigs,eigvals,svd,qr,tan,det,logspace,roll,min,mean,cumsum,cumprod,diff,vectorize,lstsq,cla,eye,xlabel,ylabel,squeeze,print},
    emphstyle=[4]\color{numpycolour},
    emph={[5]__init__,__add__,__mul__,__div__,__sub__,__call__,__getitem__,__setitem__,__eq__,__ne__,__nonzero__,__rmul__,__radd__,__repr__,__str__,__get__,__truediv__,__pow__,__name__,__future__,__all__,__main__},
    emphstyle=[5]\color{specmethodcolour},
    morekeywords=[2]{CPUQVM,PartialAmpQVM,SingleAmpQVM,MPSQVM,NoiseQVM,GPUQVM},
    morekeywords=[3]{init_qvm,directly_run,run_with_configuration,prob_run_dict,prob_run_tuple_list,prob_run_list,QProg,QVec,QCircuit,set_val,cAlloc,qAlloc_many,cAlloc_many,convert_qprog_to_qasm,convert_qprog_to_quil,convert_qprog_to_originir ,var,VariationalQuantumCircuit,init_quantum_machine,draw_qprog,init_state,aggregate_operations，Grover,build_HHL_circuit,create_if_prog,create_while_prog,Grover,Fusion,aggregate_operations,ldd_decompose,unitary2basis,OBMT_mapping, set_dagger,set_noise_model},
    morekeywords=[4]{CU,I, Z, X, Y, CNOT, CZ, T, H, S, Z,  init, measure_all, RX, RZ, CR,U3,RY,MEASURE,Measure,VariationalQuantumGate_H,VariationalQuantumGate_RX,VariationalQuantumGate_RY,VariationalQuantumGate_RZ,VariationalQuantumGate_CZ,VariationalQuantumGate_CNOT,dagger,control,assign},
    morekeywords=[5]{qubit, q,qubits,cbit,cbits,PI},
    morekeywords=[6]{DEPHASING_KRAUS_OPERATOR,BITFLIP_KRAUS_OPERATOR,DAMPING_KRAUS_OPERATOR,  set_readout_error, set_rotation_error, set_rotation_error,DEPHASING_KRAUS_OPERATOR},
    emph={[6]assert,yield},
    emphstyle=[6]\color{keywordcolour}\bfseries,
    emph={[7]range},
    emphstyle={[7]\color{keywordcolour}\bfseries},
    literate=*%
    {:}{{\literatecolour:}}{1}%
    {=}{{\literatecolour=}}{1}%
    {-}{{\literatecolour-}}{1}%
    {+}{{\literatecolour+}}{1}%
    {*}{{\literatecolour*}}{1}%
    {**}{{\literatecolour{**}}}2%
    {/}{{\literatecolour/}}{1}%
    {//}{{\literatecolour{//}}}2%
    {!}{{\literatecolour!}}{1}%
    {[}{{\literatecolour[}}{1}%
    {]}{{\literatecolour]}}{1}%
    {<}{{\literatecolour<}}{1}%
    {>}{{\literatecolour>}}{1}%
    {>>>}{\pythonprompt}{3}%
    ,%
    frame=trbl,
    xleftmargin=0.06\textwidth,
    linewidth=0.94\columnwidth,
    rulecolor=\color{black!40},
    numbers=left, 
    backgroundcolor=\color{white},
    breakindent=.5\textwidth,frame=single,breaklines=true%
}

\lstdefinestyle{PythonStyle}{
  language=Python,
  linewidth=0.94\columnwidth,
  xleftmargin=0.06\columnwidth,
  basicstyle=\linespread{1.0}\scriptsize\ttfamily, 
  captionpos=t, 
  extendedchars=true, 
  tabsize=2, 
  columns=fixed, 
  keepspaces=true, 
  breaklines=true, 
  frame=trbl, 
  rulecolor=\color{black!40},
  numbers=left, 
    commentstyle=\color{codegreen}, 
    keywordstyle=\color{eclipseGreen}, 
    keywordstyle=[2]\bfseries\color{yellow}, 
    keywordstyle=[3]\color{FFBE7A}, 
    keywordstyle=[4]\bfseries\color{FA7F6F}, 
    keywordstyle=[5]\color{0067D2}, 
    keywordstyle=[6]\color{eclipsePurple}, 
}
\lstnewenvironment{Python}[1][]{\lstset{style=PythonStyle}}{}

\begin{document}
\title{QPanda: high-performance quantum computing framework for multiple application scenarios}
\author{Menghan Dou}
\author{Tianrui Zou}
\author{Yuan Fang}
\author{Jing Wang}
\author{Dongyi Zhao}
\author{Lei Yu}
\author{Boying Chen}
\author{Wenbo Guo}
\author{Ye Li}
\affiliation{%
  \institution{Origin Quantum Computing Company Limited}
  \institution{Anhui Engineering Research Center of Quantum Computing}
  \city{hefei}
  \postcode{230026}
  \country{China}
}
\author{Zhaoyun Chen}
\affiliation{%
  \institution{Institute of Artificial Intelligence( Hefei Comprehensive National Science Center)}
  \city{hefei}
  \postcode{230026}
  \country{China}
}
\author{Guoping Guo}
\affiliation{%
  \institution{CAS Key Laboratory of Quantum Information (University of Science and Technology of China)}
  \city{hefei}
  \postcode{230026}
  \country{China}
}

\begin{abstract}
With the birth of Noisy Intermediate Scale Quantum (NISQ) devices and the verification of "quantum supremacy" in random number sampling and boson sampling, more and more fields hope to use quantum computers to solve specific problems, such as aerodynamic design, route allocation, financial option prediction, quantum chemical simulation to find new materials, and the challenge of quantum cryptography to automotive industry security. However, these fields still need to constantly explore quantum algorithms that adapt to the current NISQ machine, so a quantum programming framework that can face multi-scenarios and application needs is required. Therefore, this paper proposes QPanda, an application scenario-oriented quantum programming framework with high-performance simulation. Such as designing quantum chemical simulation algorithms based on it to explore new materials, building a quantum machine learning framework to serve finance, etc. This framework implements high-performance simulation of quantum circuits, a configuration of the fusion processing backend of quantum computers and supercomputers, and compilation and optimization methods of quantum programs for NISQ machines. Finally, the experiment shows that quantum jobs can be executed with high fidelity on the quantum processor using quantum circuit compile and optimized interface and have better simulation performance.\par
\end{abstract}
\maketitle

%

\section{Introduction}
%
%
%
%
With the development of quantum computing research, it has been transitioning from research to industrialization\cite{national2019quantum}. It is hoped to significantly improve optimization\cite{farhi2014quantum}, machine learning\cite{havlivcek2019supervised}, simulation\cite{kandala2017hardware}, and security problems\cite{gisin2002quantum}, to solve the problems encountered by existing high-performance computing systems. Applying these problems in academia and industry is a headache\cite{bayerstadler2021industry}. For example, the automotive\cite{burkacky2020will} industry's complex design, manufacturing, logistics, and financial challenges\cite{bouland2020prospects} are promising candidates based on quantum optimization\cite{farhi2014quantum} and machine learning\cite{havlivcek2019supervised} methods. Quantum chemical simulation is expected to enhance the material research process, such as battery cell chemistry, and quantum cryptography is wished to challenge the security currently dominated by Rivest–Shamir–Adleman(RSA)\cite{boneh1999twenty} and elliptic curve cryptography\cite{bos2014elliptic}.\par
Although Google's random numbers sample\cite{arute2019quantum} and Chinese researchers have created a quantum device to sample bosons\cite{zhong2020quantum}, displaying "quantum supremacy" in a problem that cannot be solved in a limited time on classical hardware. However, what quantum algorithm will bring real advantages needs to be clarified. Therefore, the quantum programming framework is becoming more and more critical for the study of quantum algorithms and the application of solving industrial requirements.\par
Current quantum computing programming frameworks usually focus on the interaction with QPU and the high performance of the quantum simulation~\cite{10.1145/3483528,10.1145/3505636}. Each framework must adapt to the development of quantum programs in multiple application scenarios. For example, most quantum computers with NISQ devices require classical quantum hybridization to achieve quantum advantage\cite{endo2021hybrid}, and supercomputers also play an essential role in quantum computing. For example, when dealing with large-scale quantum computing tasks, the quantum device does not meet the computing requirements, so the circuits need to be split with a considerable cost, which needs to be handled by supercomputers\cite{suchara2018hybrid}.\par
To meet the above requirements, we propose a quantum programming framework QPanda that supports multi-scenario applications, including:\par 
\begin{itemize}
    \item quantum finance
    \item material chemistry simulation
    \item quantum fluid mechanics
    \item  quantum machine learning
\end{itemize}\par 
It also can use supercomputers and quantum computing to solve hybrid computing problems, such as jointly as\par 
\begin{itemize}
    \item quantum error mitigation
    \item variational quantum circuits
\end{itemize}\par
At the same time, it supports the robust quantum algorithm design, enriches the single noise of QPU, and provides a controllable noise model. Provide a more comfortable quantum programming design, such as classical simulation with optimization of CPU and GPU, the most convenient tools for use, etc. Finally, to provide more efficient QPU interaction, we use quantum compilation optimization technology to assist QPU in executing high-fidelity metric subprograms.\par
To prove the high performance of QPanda, we list corresponding comparison experiments, and experimental results verify that when QPanda simulates random circuits, whether having a better performance in single-threaded or multi-threaded. And then, we list the result of the circuit compiled to QPU execute, which has high fidelity.\par 
This paper includes the following sections. Section ~\ref{overview} outlines the design principles and structure of QPanda. Section~\ref{Implementation} introduces the high-performance scheme in the QPanda framework, supercomputer-based quantum-classical hybrid algorithm support, partitioning circuit, classical simulation method, convenient tools of QPanda, and the implementation of various simulations. Section~\ref{experiment} presents the application experiment of QPanda and the performance comparison experiment with the current mainstream quantum simulation framework. Section~\ref{conclusion} is a conclusion and future work.\par

\section{Overview}\label{overview}
At present, quantum computers are limited to minor problems that could be more practical. However, promising hybrid algorithms for these NISQ systems have emerged using both classical and quantum hardware. Although most of the current algorithms in the field of quantum computing are designed similarly, they address different problems. It includes simulating the chemical properties of small molecules, solving optimization problems, performing machine learning tasks, or solving differential equations. The implementation of these application algorithms requires an efficient quantum programming framework. For example, researchers do not care about the underlying logic in designing algorithms but only about the corresponding results of algorithms. Therefore, a framework compatible with high-level quantum languages can enable researchers to focus more on algorithm design. In addition, the robust design of quantum algorithms often requires an adjustable noise for testing, and the stable noise of QPU in the period often needs to meet the requirements. Of course, what is more critical is the performance of the framework. For example, the performance of simulation and the performance of interaction with QPU are all considered by researchers.\par
\begin{figure*}[htbp]
    \centering
    \includegraphics[width=13cm]{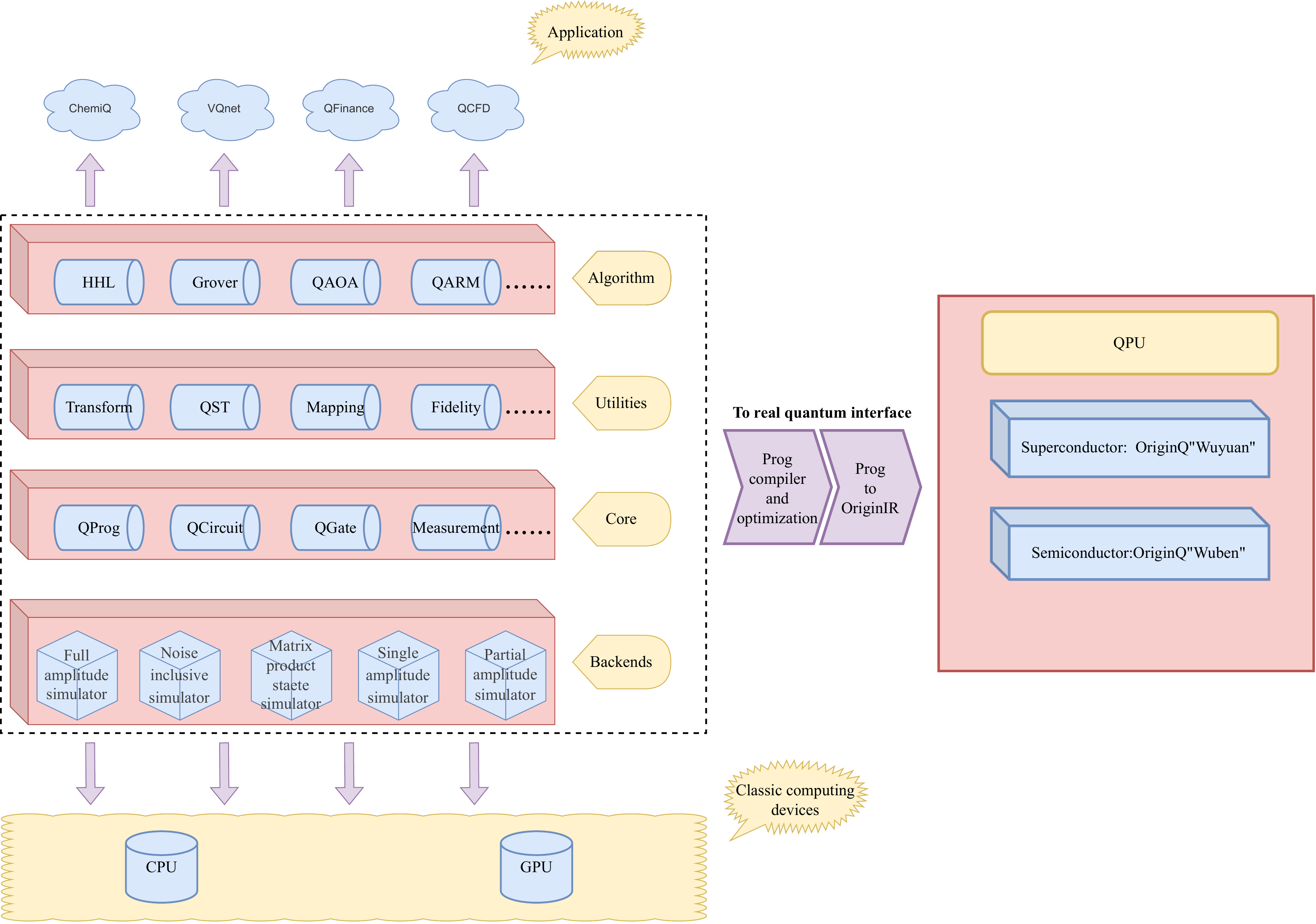}
    \caption{QPanda framework}
    \label{fig:2.1}
\end{figure*}
\subsection{Design Principles}\label{principles}
QPanda's design principles are applicated to multi-application scenarios and high performance. It is used to promote the transition of quantum computing from research to industrialization.\par
\textbf{High-performance}: Although there are hundreds of qubits of quantum computers\cite{chow2021ibm}, due to the small quantum volume and noise calibration problems, classical simulation has become indispensable. High-performance computing has always been the goal of people, and quantum computing is nothing more than that. Therefore, high-performance simulation has become one of the original intentions of QPanda.\par 
It is well known that the simulation time of quantum circuits increases exponentially with the number of qubits\cite{arute2019quantum}. To reduce simulation time, QPanda processes quantum circuits in parallel on single-core or multi-core CPUs through OpenMP, uses GPU (graphics processing unit) to optimize calculation, and optimizes quantum circuits to reduce quantum logic gates before simulation. Compared with the unoptimized quantum circuit simulation, these methods can achieve a performance improvement of 5 times.\par
\textbf{Multi-scenario application support}: To support the above multi-scenario research and development needs, QPanda introduced tools such as variational quantum circuit, Hamiltonian decomposition, and unitary matrix decomposition to accelerate algorithm design. It provides a quantum supercomputer fusion backend to support quantum error mitigation, variational quantum circuits, other hybrid algorithms, and multiple simulation backend processes with different requirements. As well as the quantum compilation optimization process, quantum programs can run on QPU with high fidelity.\par
\subsubsection{A variety of convenient tools}
To meet various requirements in quantum computing research, QPanda provides various convenient tools for researchers.\par
\begin{itemize}
    \item Since the circuit consists of fundamental quantum gates, QPanda provides many standard gates for researchers. To simplify the usage of control gates, especially the multi-control gates, QPanda provides a convenient representation of multi-control gates, that is, by setting control qubits through \Colorbox{bkgd}{.control} after a target gate.
    \item Quantum machine learning is significant research in quantum computing, and QPanda also provides many basic algorithms, which effectively simplify the development time of researchers. For example, variable quantum algorithm interfaces, Hamiltonian, and other quantum algorithms for researchers.
    \item QPanda provides \Colorbox{bkgd}{Qif} and\Colorbox{bkgd}{QWhile} interfaces to implement  While and if classical operations. It also supports various classical operators, such as and, or, addition, multiplication, and other operators, simplifying the process of quantum programs.
    \item To realize the reusability of quantum circuits, QPanda separates quantum circuits from quantum programs so that researchers can freely design circuits. In contrast, circuit measurement operation is embedded in quantum programs, which significantly broadens the freedom of quantum programming.
    \item To obtain the quantum state of quantum circuits, QPanda provides quantum state vectors, probability, expectation, and density matrices tools to help researchers observe the circuit.
\end{itemize}

\subsubsection{Quantum-supercomputer fusion backend}
In the NISQ era, all the qubits we control are noise qubits, and the gate error may be $10^{-3}$ or lower. Although the NISQ computer is not general-purpose, we can solve some computing tasks by combining quantum computers and supercomputers. Such as quantum error mitigation technology for error compensation through experimental post-processing data, machine learning algorithm based on the variational quantum circuit, quantum circuit segmentation, etc.\par
\subsubsection{Multiple simulation backend}\par
We not only propose four kinds of noiseless simulation backend, namely full amplitude, partial amplitude, single amplitude, and matrix product state (MPS) simulators, to meet different theoretical simulation requirements. It also provides an adjustable noise simulator for the current research on the robustness of quantum algorithms, which is used to solve the single noise demand of QPU in the same period.\par
\subsubsection{Efficiently compile quantum programs}\par
Limited by the topology, noise, and basis quantum gates of NISQ superconducting quantum computers, we have to optimize the compilation of quantum programs to make the compiled quantum circuit shorter, that is, higher fidelity. Therefore, we provide a multi-level quantum program compilation and optimization framework to ensure the high-fidelity operation of quantum programs.\par
\begin{figure*}
    \centering
    \includegraphics[width=13cm]{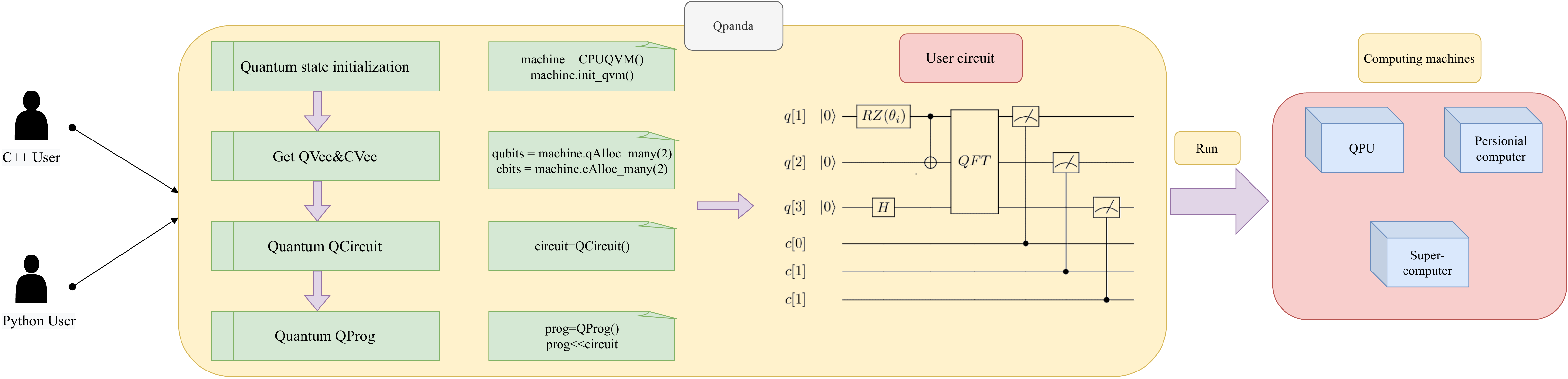}
    \caption{The overview of using QPanda}
    \label{fig:2.2}
\end{figure*}

\subsection{Structure of QPanda}\label{structures}
As shown in Figure \ref{fig:2.1}, the QPanda quantum programming framework is divided into the Application, Algorithm, Utilities, Core, and Computing backend. Below, we will describe each part in detail.\par
\textbf{Application}
This layer includes quantum chemistry (ChemiQ\cite{wang2021chemiq}), quantum machine learning (VQNet\cite{chen2019vqnet}), quantum finance applications (QFinance), quantum computational fluid dynamics (QCFD).\par
\textbf{Algorithm}
The algorithm layer mainly includes Hamiltonian Simulation\cite{sheldon2016procedure}, MaxCutProblem Generator\cite{guerreschi2019qaoa}, Pauli operator\cite{schiller1999pauli} and Fermion operator\cite{aguilar2011effective}, Variational quantum algorithm\cite{wecker2015progress}, etc. The algorithm library and application mainly include quantum operators (add, subtract, multiply, etc.), BV algorithm\cite{bernstein1997quantum,du2001implementation}, DJ algorithm\cite{deutsch1992rapid,cleve1998quantum}, Grover algorithm\cite{grover1996fast}, HHL algorithm\cite{harrow2009quantum}, QAOA algorithm\cite{farhi2014quantum}, QARM algorithm\cite{lei2005improvements}, QITE algorithm\cite{mcardle2019variational, yeter2020practical, yeter2021scattering}, Shor algorithm\cite{shor1999polynomial, gidney2021factor}, etc.\par
In Code~\ref{code:QPanda_HHL},~\ref{code:QPanda_Grover}, we display a little case of the HHL and Grover algorithms utilities and provide the highly integrated modular interface to realize algorithm functions.\par
\begin{lstfloat}
\begin{lstlisting}[
  language=Python,
  style=PythonStyle,
  caption=QPanda HHL example.,
  label=code:QPanda_HHL,
]
# Init quantum simulator
machine = CPUQVM()
machine.init_qvm()

# Building HHL circuit  
build_HHL_circuit([1,0,0,1],[0.6,0.8],machine)
\end{lstlisting}

\begin{lstlisting}[
  language=Python,
  style=PythonStyle,
  caption=QPanda Grover example.,
  label=code:QPanda_Grover,
]
# Init quantum simulator and allocation classical register 
machine = CPUQVM()
machine.init_qvm()   
x = machine.cAlloc()  
  
# searching space   
data = [8, 7, 6, 0, 6, 3, 6, 4, 21, 15, 11, 11, 3, 9, 7]  
measure_qubits = machine.QVec()

# grover circuit module
grover_circuit = Grover(data, x == 6, machine, measure_qubits, 1)  
    
\end{lstlisting}
\end{lstfloat}

\textbf{Utilities}
It mainly includes quantum tools and transforms. The primary function of the devices is to obtain relevant information about quantum programs, such as judging whether quantum gates are interchangeable and compatible, obtaining quantum program layering, the number of gates, etc. The quantum program transforms textualizing the quantum program, changing the target quantum language into other quantum programs to adapt the target quantum computing platform.\par
\textbf{Core}
The core layer is mainly divided into the quantum simulation backend and the quantum circuit modules. The quantum simulation backend module uses various quantum simulation methods to build a simulation backend for different requirements\ref{fig:2.3}. The quantum circuit module provides a way to construct quantum programs. It includes \Colorbox{bkgd}{QProg}, \Colorbox{bkgd}{QCircuit}, \Colorbox{bkgd}{QMeasure}, \Colorbox{bkgd}{QGate}, \Colorbox{bkgd}{QIf}, \Colorbox{bkgd}{QWhile}, \Colorbox{bkgd}{ClassicalProg} seven types. The code example as followed in Code~\ref{code:QPanda_backend},~\ref{code:QPanda_Qif},~\ref{code:QPanda_QWhile}.\par
\begin{lstfloat}[bt]
\begin{lstlisting}[
  style=PythonStyle,
  numbers=left,
  linewidth=0.94\columnwidth,
  xleftmargin=0.06\columnwidth,
  basicstyle=\linespread{1}\scriptsize\ttfamily,
  caption=QPanda simulation backend example.,
  label=code:QPanda_backend,
]
# Apply for full-amplitude quantum simulator 
machine = CPUQVM()

# Apply for partial-amplitude quantum simulator 
machine = PartialAmpQVM()

# Apply for single-amplitude quantum simulator
machine = SingleAmpQVM() 

# Apply for tensor network quantum simulator
machine = MPSQVM()

# Apply for noise inclusive quantum simulator
machine = NoiseQVM()


 
\end{lstlisting}
\end{lstfloat}

\begin{lstfloat}
\begin{lstlisting}[
  style=PythonStyle,
  numbers=left,
  linewidth=0.94\columnwidth,
  xleftmargin=0.06\columnwidth,
  basicstyle=\linespread{1}\scriptsize\ttfamily,
  caption=QPanda Qif example.,
  label=code:QPanda_Qif,
]
# Init quantum simulator and allocation quantum and classical register
machine = CPUQVM()
machine.init_qvm()
qubits = machine.qAlloc_many(3)
cbits = machine.cAlloc_many(3)
	
# Set the initial value of the classic bit
cbits[0].set_val(0)
cbits[1].set_val(3)

prog = QProg()
branch_true = QProg()
branch_false = QProg()

# Building true and false prog
branch_true << H(qubits[0])<< H(qubits[1]) << H(qubits[2])
branch_false << H(qubits[0]) << CNOT(qubits[0], qubits[1]) << CNOT(qubits[1], qubits[2])

# Building QIf
qif = create_if_prog(cbits[0] > cbits[1], branch_true, branch_false)
\end{lstlisting}

\begin{lstlisting}[
  style=PythonStyle,
  numbers=left,
  linewidth=0.94\columnwidth,
  xleftmargin=0.06\columnwidth,
  basicstyle=\linespread{1}\scriptsize\ttfamily,
  caption=QPanda QWhile example.,
  label=code:QPanda_QWhile,
]
# Init quantum simulator and allocation quantum and classical register
machine = CPUQVM()
machine.init_qvm()
qubits = machine.qAlloc_many(3)
cbits = machine.cAlloc_many(3)
prog_while = QProg()

# Set the initial value of the classic bit
cbits[0].set_val(0)
cbits[1].set_val(1)

# Building QWhile program
prog_while << H(qubits[0]) << H(qubits[1])<< H(qubits[2])<< assign(cbits[0], cbits[0] + 1) << Measure(qubits[1], cbits[1])
qwhile = create_while_prog(cbits[1], prog_while)
\end{lstlisting}
\end{lstfloat}
\textbf{Computing backend}
The computing backend is the unit used to complete the quantum program calculation. It can be accelerated by CPU or GPU. In addition, QPanda uses Eigen\cite{guennebaud2010eigen} to calculate matrices, uses pybind11\cite{jakob2017pybind11}  to export functions and classes from C++ to Python, and can be tested with GoogleTest\cite{GoogleTest} and Pytest\cite{krekel2004pytest}. Researchers can also use the QPU service of OriginQ company through QPanda and direct experience of the quantum advantage\cite{Cloud}. The ways to know the most details of QPanda are the official website documentation\footnote{The tutorial of pyqpanda can be found at:
\textcolor{blue}{
\url{https://pyqpanda-tutorial-en.readthedocs.io/en/latest/}}.} or the source code from Github\footnote{The code of QPanda can be found at:
\textcolor{blue}{
\url{https://github.com/OriginQ/QPanda-2}}.}. Figure \ref{fig:2.2} shows the researchers used QPanda to simulate the quantum circuit procedure.\par
\section{Implementation of QPanda}\label{Implementation}
\subsection{High-performance simulation}\label{high_performance}
QPanda uses OpenMP, GPU, and circuit optimization methods to solve the performance bottleneck problem in quantum circuit simulation. Because quantum circuit computation is a simple iteration of quantum gates, the time of circuit simulation can be roughly divided into the time cost of calling functions, computation process, and memory operation. Some factors include calling functions, parallel functions, and the GPU kernel through the python interface, which can cause additional time overhead. CPU computing power is determined by the number of operations processed per second, called floating point operations per second (FLOPS). In the simulation of quantum circuits, there are many high-performance optimization ideas, such as the parallelism of quantum gate computation, complex operation in quantum states or gates, and optimal merging of logic gates in quantum circuits. Based on the above optimization ideas, QPanda uses open multiprocessing (OpenMP) for CPU parallel processing and GPU for multi-core computing. It uses single instruction multiple data (SIMD) technology to load complex quantum states from the CPU cache or RAM to CPU registers. GPU quantum circuit fusion scheme is used for circuit optimization and consolidation.\par
\subsubsection{OpenMP multi-thread optimization}\label{openmp}\par
Existing CPUs contain multiple processing cores, and the parallel execution of iterations via these cores in functions can reduce computation time. A single CPU core has a certain workload. Here, QPanda uses OpenMP parallel execution functions to add threads for parallel execution to improve the computing efficiency when the core FLOPS reaches the performance bottleneck.\cite{dagum1998openmp}. The number of threads used in these functions can be determined by the \Colorbox{bkgd}{OMP\_NUM\_threads}. In QPanda, when the backend is used to execute large depth or wide quantum circuits, OpenMP will increase memory consumption and time costs because the amount of computation is too small to reach the payload of a single core. To maximize the workload of each kernel and minimize the multithreading overhead, we set the threshold to control OpenMP. The quantum gates and qubits determine this threshold.\par
\subsubsection{GPU optimization}\label{GPU}\par
Compared to the CPU, GPU has multiple cores. Although the CPU has many advantages, such as logic control and serial computing, it's less efficient when dealing with vast and repeatable arithmetic jobs. However, the GPU's massively parallel processing uses a particular parallel computing algorithm to improve efficiency and quantum simulation tally with this feature of the GPU. For example, computing the evolved states via the quantum gate is a vast and repetitive job in the full-amplitude simulation method. \par
Due to the GPU having a memory hierarchy, GPU optimizes simulation and emphasizes GPU memory allocation and data transmission from CPU to GPU. In NVIDIA GPU, memory is classified into six categories: global, local, shared, registers, constant, and texture. These memory types are distributed on RAM or GPU and have different memory sizes and information interaction schemes. QPanda adaptive allocates the required memory in the corresponding location according to the simulated qubits and gates to avoid additional memory allocation or access during the arithmetical operation. For example, use shared memory to store the gates when disposing of small computing jobs, improve the efficiency of each gate copy and modify which from the CPU to the GPU during the arithmetical operation. In addition, QPanda will select global memory to replace shared memory when dealing with large computing jobs. Finally, researchers via the \Colorbox{bkgd}{GPUQVM}interface to optimize their circuit, the example is in Code~\ref{code:QPanda_gpuqvm}. \par
\begin{lstfloat}[bt]
\begin{lstlisting}[
  style=PythonStyle,
  numbers=left,
  linewidth=0.94\columnwidth,
  xleftmargin=0.06\columnwidth,
  basicstyle=\linespread{1}\scriptsize\ttfamily,
  caption=QPanda GPU optimization example.,
  label=code:QPanda_gpuqvm,
]
# Apply for GPU optimization simulator 
machine = GPUQVM()
\end{lstlisting}
\end{lstfloat}
\subsubsection{Quantum circuit fusion}\label{GPU}
Since there may exist redundant circuits in a quantum program, the optimized quantum circuit executes the quantum program more efficiently without modifying the quantum algorithm.\par
In most cases, the quantum circuits to be optimized have complex implementations, which are mainly manifested in huge qubits and gates. As shown in the figure\ref{fig:optimize}, the dotted part of the quantum circuit represents the local circuit to be optimized. After the local circuit optimization is completed, the algorithm will automatically update the optimized position, three unitary matrix blocks in the right half in the figure\ref{fig:optimize} layer 2. There are several ways to optimize the circuit. In Code~\ref{code:QPanda_opt}, researchers can call the \Colorbox{bkgd}{aggregate\_operations} member function in the \Colorbox{bkgd}{Fusion} class to optimize their circuit. \par
\begin{lstfloat}[bt]
\begin{lstlisting}[
  style=PythonStyle,
  numbers=left,
  linewidth=0.94\columnwidth,
  xleftmargin=0.06\columnwidth,
  basicstyle=\linespread{1}\scriptsize\ttfamily,
  caption=QPanda circuit fusion and optimization example.,
  label=code:QPanda_opt,
]
# Init quantum simulator and allocation quantum and classical register
machine = CPUQVM()
machine.init_qvm()  
qubits = machine.qAlloc_many(4)  
cbits = machine.cAlloc_many(4)  
prog = QProg()  
  
# Building a quantum program 
prog << X(qubits[0]) << Y(qubits[1])\  
          << H(qubits[2]) << RX(qubits[3], PI)\  
          << CNOT(qubits[0],qubits[1])  
  
# Circuit fusion and optimization  
fuser = Fusion()  
fuser.aggregate_operations(prog, machine)
\end{lstlisting}
\end{lstfloat}
\begin{figure*}[htbp]
    \centering
    \includegraphics[width=13cm]{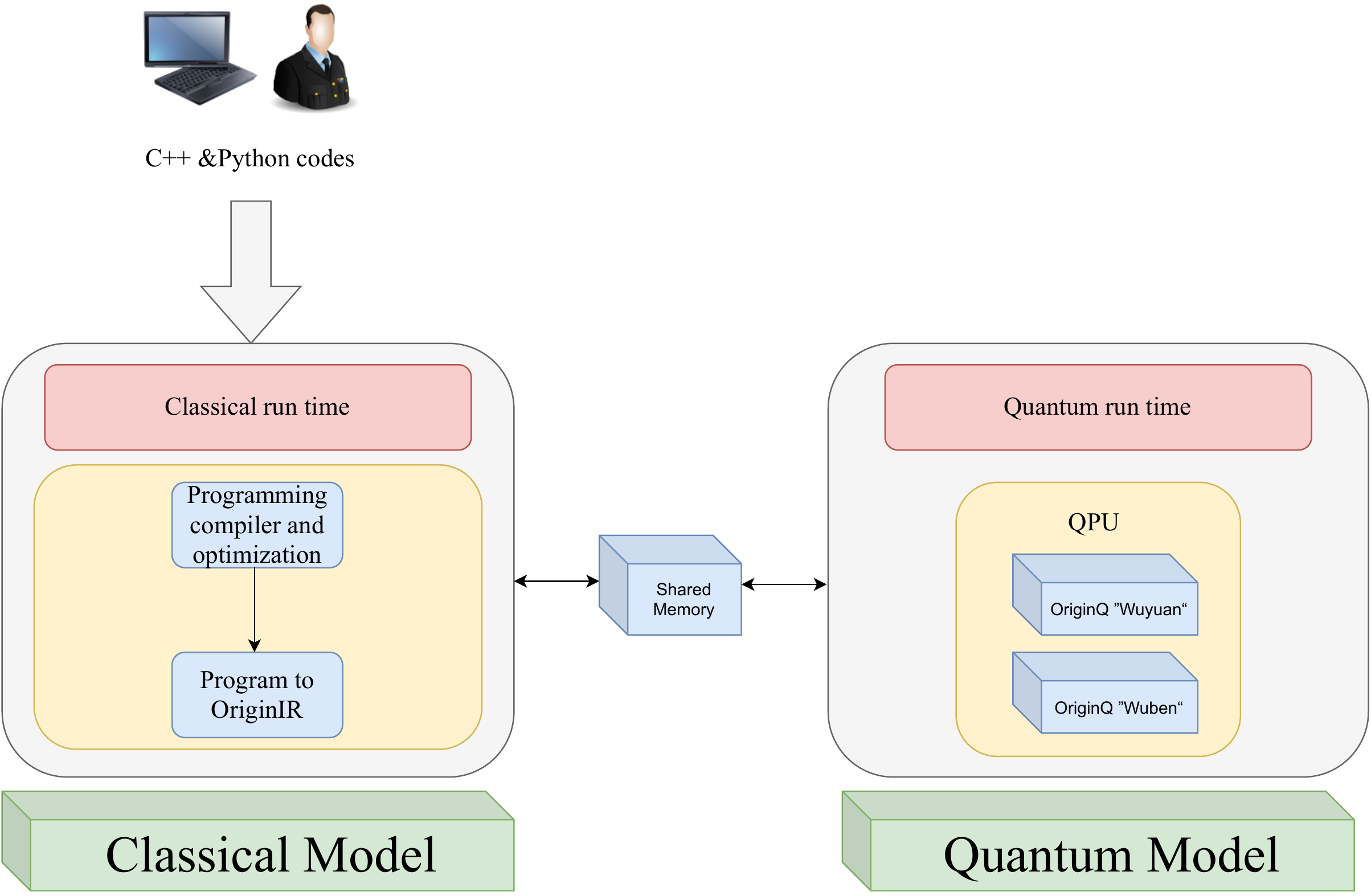}
    \caption{Overview of the QPanda running framework}
    \label{fig:3.2}
\end{figure*}
\subsection{Quantum-supercomputer fusion backend}
\subsubsection{Simulate huge circuits}
To deal with the limitation of hardware facilities, QPanda proposes a diversified deployment scheme. This scheme solves the hardware limitation of personal computers when simulating high qubits, which use the bi-level parallel method on the supercomputer to improve efficiency.\par
The first level is the parallelism in space, splitting a large computing job and evenly allocating to N nodes, then serial computing on these N nodes. The next level's serial computing jobs are divided into multiple threads or processes. Each node hardly participates in the computing process of the core when performing computing tasks, and is mainly responsible for I/O operations, task scheduling and assignment of computing cores, and data communication with other processes. The next level divides the serial jobs of each node into multi-thread or multi-process execution based on the first level. Each node is mainly responsible for I/O operations, scheduling, assignment of computing cores, and data transmission with other processes, multiple cores of the node handles the computing tasks.\par
\subsubsection{Quantum circuit partition}\par
We use classical computers to decompose quantum circuits and run sub-circuits in quantum computers to solve the problem that the NISQ computer is short of quantum bits and cannot deal with large-scale quantum circuits. These tools need to achieve a long list of conflicting goals. To reduce the cost, we designed a circuit compiler splitter to exert the computing power of supercomputers. At the same time, the compiler can also identify which subcircuits can be effectively simulated on a classical computer to reduce the pressure on quantum devices.\par
\subsubsection{Support hybrid quantum-classical algorithm}\par
Because NISQ devices can only use relatively external circuits on a limited number of qubits, traditional quantum algorithms may not be implemented on NISQ devices. Therefore, it is necessary to consider hybrid quantum algorithms to give full play to quantum advantages.\par
\subsection{QPU executes QPanda quantum programs }\label{QPU}
As mentioned above, executing a quantum algorithm on the QPU to verify quantum advantage and adjust is an indispensable step in quantum algorithms. So, QPanda provides an interface for QPU to execute the quantum programs. As shown in Figure \ref{fig:3.2}, the compilation and execution process of quantum programs in QPanda is described. We divide the quantum program into two modules for execution: the classical and quantum modules. In the classical module, the quantum program is compiled, optimized, and then translated into OriginIR for submission to the QPU; the classical and quantum modules are transmitted through shared memory.\par
\subsubsection{QPanda compilation and optimization}\label{QC}\par
\begin{figure}
    \centering
    \includegraphics[width=7cm]{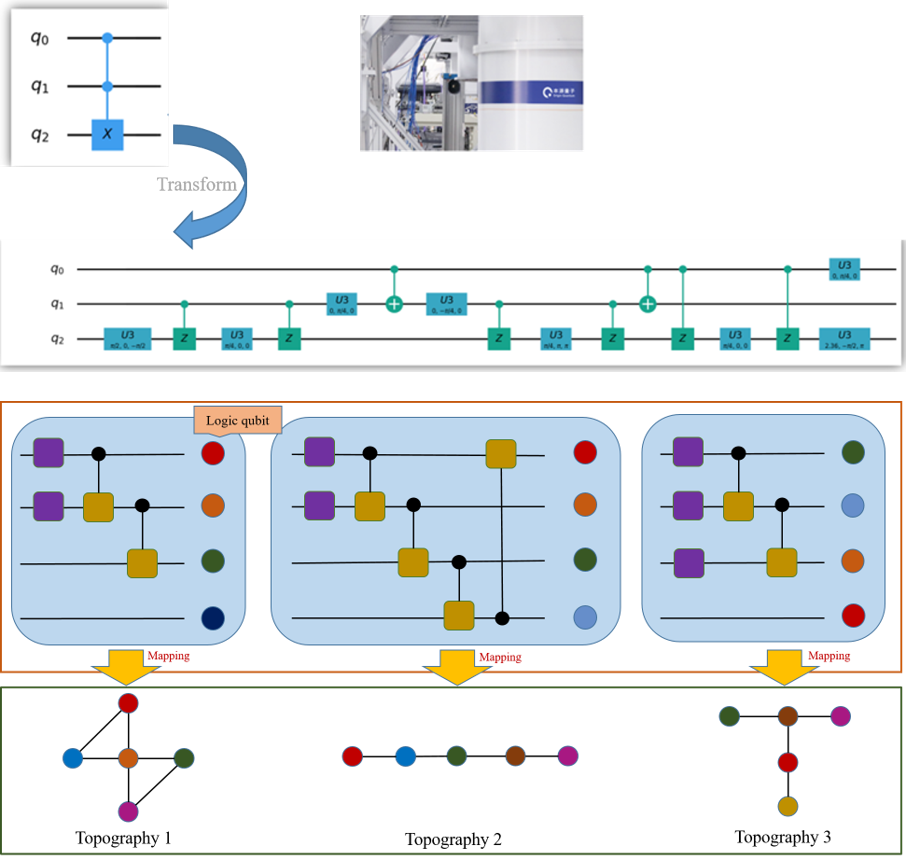}
    \caption{Quantum program compilation framework}
    \label{fig:qpc}
\end{figure}

Since QPU is not a classical computer, quantum programs cannot execute directly on QPU. Therefore, it is necessary to compile and optimize the quantum program to submit, and the process is shown in Figure \ref{fig:qpc}. First, decompose the control gates into a set of gates that conform to the QPU rules. Then replaces the irregular gates and optimizes the circuit to reduce the number of gates. Finally, it uses the mapping scheme to map the virtual qubits to the physical properties of the chip Topology.\par
\begin{figure}
    \centering
    \includegraphics[width=7cm]{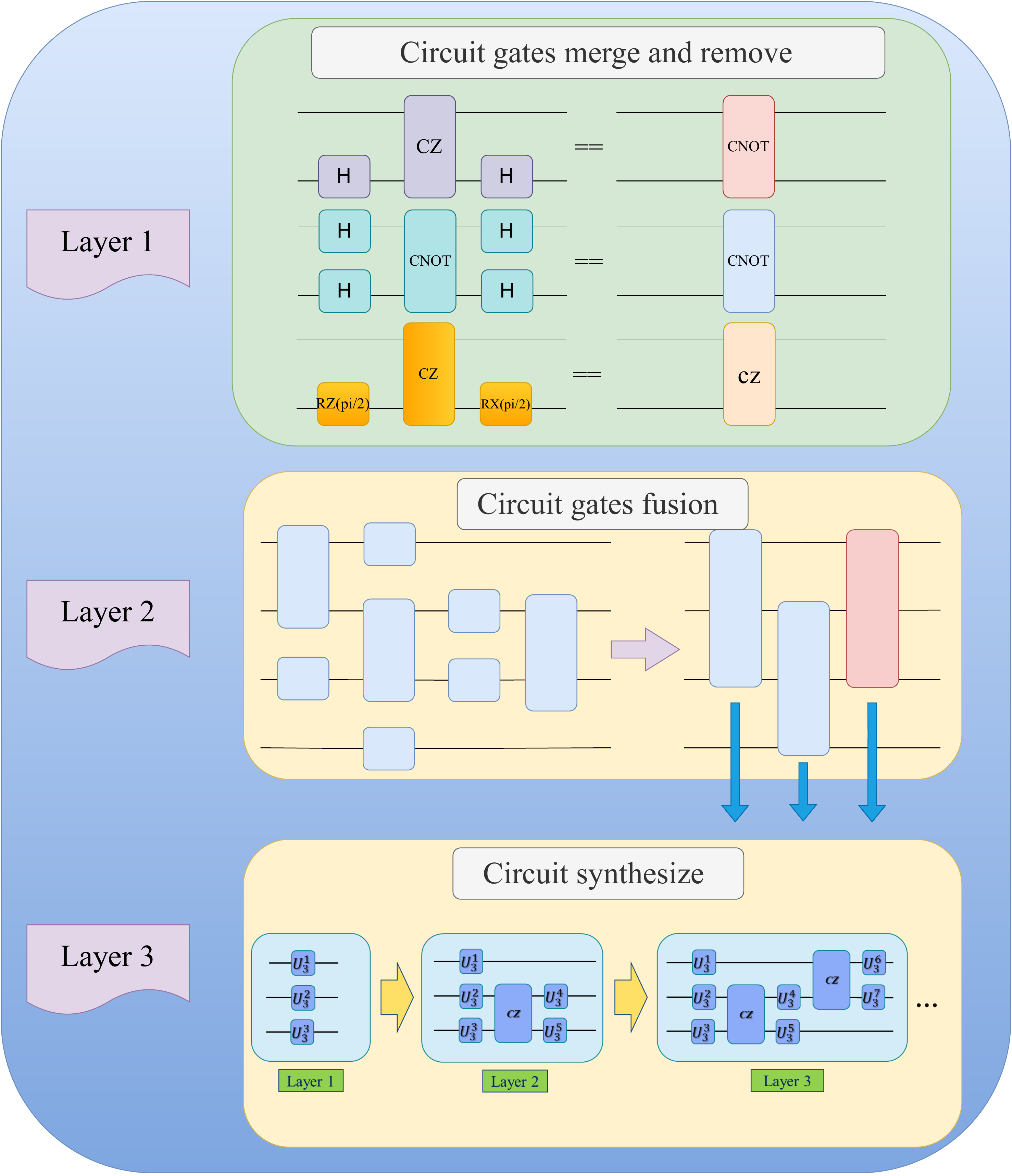}
    \caption{Quantum circuit optimize framework}
    \label{fig:optimize}
\end{figure}
\textbf{1.Multi-control gates decomposition}\par
Many basic gates cannot be directly compiled to the target quantum chip. Therefore, to compile the target quantum program to chip, the multi-control gates in the circuit need to be decomposed into multiple basis gates with the \colorbox{bkgd}{ldd\_decompose} in Code~\ref{code:QPanda_decompose}. For example, the control gate will be transformed into $U_3$ and CZ gate in the "Wuyuan" chip provided by OriginQ.\par
\begin{lstfloat}[bt]
\begin{lstlisting}[
  style=PythonStyle,
  numbers=left,
  linewidth=0.94\columnwidth,
  xleftmargin=0.06\columnwidth,
  basicstyle=\linespread{1}\scriptsize\ttfamily,
  caption=Multi-control gates decomposition example.,
  label=code:QPanda_decompose,
]
# multi control gates decomposition for U in SU(2)
new_prog = ldd_decompose(prog)
\end{lstlisting}
\end{lstfloat}
\textbf{2.Basis unitary gate transform}\par
In Code~\ref{code:QPanda_basis}, due to the basis gates supported by different quantum chips not exactly, QPanda will check and translate the gates based on the quantum chip when a quantum program executes on QPU with the \colorbox{bkgd}{unitary2basis}. This ensures that the program can run correctly on the target chip.\par
\begin{lstfloat}[bt]
\begin{lstlisting}[
  style=PythonStyle,
  numbers=left,
  linewidth=0.94\columnwidth,
  xleftmargin=0.06\columnwidth,
  basicstyle=\linespread{1}\scriptsize\ttfamily,
  caption=Basic unitary gate transform example.,
  label=code:QPanda_basis,
]
# Basis unitary gate transform
# basis_gates is the basis unitary gates based on the target chip
new_prog = unitary2basis(prog, machine, basis_gates);
\end{lstlisting}
\end{lstfloat}
\textbf{3.Quantum circuit optimization}\par
The quantum job will be completed on QPU when the gates are decomposition and replaced. From Fig \ref{fig:optimize}, the transformed quantum circuit can be further optimized with three layers which ensure the entire quantum program is a more efficient operation. The first layer details as follow:\par
1. Remove useless identity pair gates\par
Without considering noise, two consecutive Hadamard gates or X gates can delete in the circuit, and the effect is equivalent to an identity gate (I gate).\par
2. Merge gates\par
The following optimal combination of gates situations: two consecutive $R_z$ gates can be merged into one $R_z$ gate, and the rotation angle of the newly synthesized $R_z$ gate is the sum of the angles of the two $R_z$ gates before the merger. Similarly, $R_x$, $R_y$ gates can also be merged. In addition, for the gates that can be combined, a plurality of consecutive gates can be combined instead of being limited to combining two consecutive gates. \par
Then, The simulation of multi-qubit gates in quantum circuits is generally divided into two cases. The first case is to decompose the multi-qubit gates. As mentioned above, the multi-qubit gates cannot be directly compiled and executed in QPU, so we use the circuit's DAG to obtain the fusion gates. Then synthesize the fused gates with U3 and CZ ansatz, which is shown in the second and third layers of Fig\ref{fig:optimize}; the second case is that control gates are fused into a sizeable unitary matrix. 
\textbf{4.Quantum mapping}\par
The topology reflects the entanglement relationship between the chip's qubits. There may be a qubit entanglement problem that the target chip does not support when building a circuit. In particular, when a quantum program executes on different chips, mapping the virtual qubit to the physical qubit of the chip is extremely important, which ensures the quantum job process is accurate and efficient. In Code~\ref{code:QPanda_chip}, we can use \colorbox{bkgd}{OBMT\_mapping} to meet the chip.\par
\begin{lstfloat}[bt]
\begin{lstlisting}[
  style=PythonStyle,
  numbers=left,
  linewidth=0.94\columnwidth,
  xleftmargin=0.06\columnwidth,
  basicstyle=\linespread{1}\scriptsize\ttfamily,
  caption=Quantum mapping example.,
  label=code:QPanda_chip,
]
# Quantum mapping
# prog: the target prog
# quantum_machine: quantum machine
# b_optimization: whether open the optimization
# max_partial: Limits the max number of partial solutions per step, 
# there is no limit by default
# max_children: Limits the max number of candidate-solutions per double gate, 
# there is no limit by default
# config_data is a configuration file or configuration data, 
# which is the chip's performance and topological structure, such as coupling map, 
# qubits fidelity, and single or double quantum gates fidelity
OBMT_mapping(prog, machine, b_optimization, max_partial = 4294967295, max_children = 4294967295, config_data)
\end{lstlisting}
\end{lstfloat}
QPanda utilizes the opt-bmt\cite{siraichi2019qubit}mapping algorithm, which is suitable for the current NISQ situation where qubits are scarce, and the high depth of the circuit. The high fidelity of the quantum jobs mapped by opt-bmt\cite{siraichi2019qubit}. Firstly, it transforms the circuit into a Directed Acyclic Graph (DAG: Directed Acyclic Graph) and then traverses from the node with in-degree 0 in the DAG. There will be N corresponding in-degrees for a circuit with N qubits. Nodes with a degree of 0, then the entire traversal process starts from these N nodes simultaneously. In the traversal process, the topology and fidelity of the chip are used as the in-degree conditions; for one or more gates that the topology of the chip without adaptability, connect it to the adjacent qubit through the swap gate to make it satisfy the chip. \par
\subsubsection{Program to OriginIR}\label{OriginIR}
Since the mapped circuit cannot be directly executed on the QPU as a job, it needs to be translated into the assembly language (OriginIR) to submit. OriginIR is an intermediate representation of quantum programs based on QPanda, which supports various features of QPanda. In addition, its implementation method is simple, directly oriented to QPU, and can be quickly compiled into binary executable files to execute on QPU.\par

\subsubsection{Multi-simulation methods}\label{S_m}
QPanda provides a noiseless and noisy simulation backend to achieve various simulation purposes. It includes four noise-free theoretical numerical simulation backends, which can quickly verify the quantum algorithm. In addition, to satisfy researchers' research on the robustness of quantum algorithms, an adjustable noise simulation backend is provided.\par
\begin{figure*}
    \centering
    \includegraphics[width=13cm]{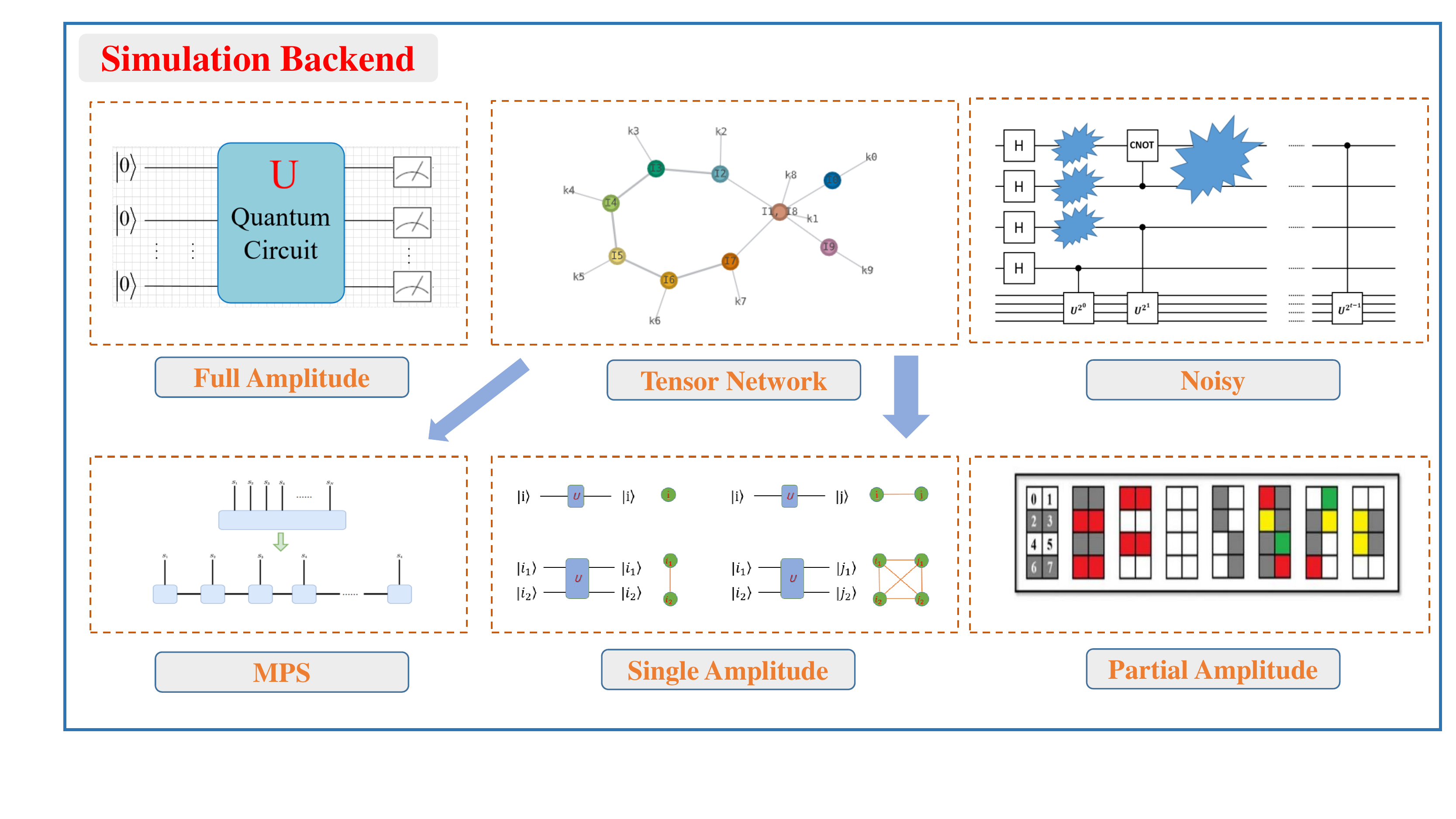}
    \caption{Simulation backends}
    \label{fig:2.3}
\end{figure*}
\textbf{1.Noise-free simulation method}\par
From figure~\ref{fig:2.3}, the provided noiseless theoretical simulation backend is usually based on two primary methods, namely the Schrodinger and Feynman methods. Schrodinger method's core is to update the quantum state represented by state vector or density matrix by applying quantum gate as a general mapping\cite{arute2019quantum}. The advantage of this method is that it simulates the quantum circuit quickly and universally, but the machine memory will limit it. QPanda proposes the full-amplitude and partial-amplitude simulation backends(\Colorbox{bkgd}{CPUQVM},\Colorbox{bkgd}{PartialAmpQVM}) based on this approach. The partial-amplitude simulation backend is designed to solve the memory-limited problem of full-amplitude simulations by transforming several CZ gates and CNOT gates into a set of single-qubit gates. This circuit is mapped onto additional subcircuits. These subcircuits consist of two blocks without any qubits entanglement between them, thus transforming the N qubits simulation problem into a set of N/2 qubits subcircuits. Another approach is to simulate quantum circuits by summing Feynman's contributions to all paths. This approach dramatically reduces the memory requirement and obtains a single amplitude of the final quantum state quickly and easily. Although the sum of Feynman's paths does not increase with qubit's width, it increases exponentially with the quantum gates\cite{arute2019quantum,markov2008simulating,alshawi2012lifetime}, QPanda also utilizes Feynman's method to design a corresponding single-amplitude simulate backend\Colorbox{bkgd}{SingleAmpQVM}. Later, to alleviate the situation that the simulation time increases exponentially with the increase of the quantum gates, QPanda helps this situation based on the tensor network\cite {alshawi2012lifetime} and designs the tensor network simulate backend\Colorbox{bkgd}{MPSQVM} so that the simulation time increases exponentially according to the treewidth of the network.\par
\textbf{2.Noise simulation method}\par
To help study the robustness of quantum algorithms under different noises, we propose a noisy quantum simulation backend\Colorbox{bkgd}{NoiseQVM} embedded with an adjustable error rate Kraus operator noise model and set Qubit fidelity. As well as multiple QPU sampling to generate the corresponding noise data set and then through machine learning to achieve approximate QPU noise training.\par
QPanda provides varity noise models with kraus operator~\cite{10.5555/2871422.2871425}:
\begin{itemize}
    \item Damping kraus operator
    \begin{equation}
    K_{1}=\left[\begin{array}{cc}
    1 & 0 \\
    0 & \sqrt{1-p}
    \end{array}\right], K_{2}=\left[\begin{array}{cc}
    0 & \sqrt{p} \\
    0 & 0
    \end{array}\right]
    \label{eq:damping}
    \end{equation}
    \item Dephasing kraus operator
    \begin{equation}
    K_{1}=\left[\begin{array}{cc}
    \sqrt{1-p} & 0 \\
    0 & \sqrt{1-p}
    \end{array}\right], K_{2}=\left[\begin{array}{cc}
    \sqrt{p} & 0 \\
    0 & -\sqrt{p}
    \end{array}\right]
    \label{eq:dephasing}
    \end{equation}
    \item Decoherence kraus operator
    \begin{equation}
    \begin{aligned}
    &P_{\text {damping }}=1-e^{-\frac{\text { tgate }}{T_{1}}} \\
    &P_{\text {dephasing }}=0.5 \times\left(1-e^{-\left(\frac{\text { tgate }}{T_{2}}-\frac{\text { tgate }}{2 T_{1}}\right)}\right)\\
    &K_{1}=K_{1_{\text {damping }}} K_{1_{\text {dephasing }}} K_{2}=K_{1 \text { damping }} K_{2 \text { dephasing }}\\
    &K_{3}=K_{2 \text { damping }} K_{1 \text { dephasing }} K_{4}=K_{2 \text { damping }} K_{2 \text { dephasing }}
    \end{aligned}
    \end{equation}
    Where $K_{\text {damping }}$ is Eq.~\ref{eq:damping}, $K_{\text {dephasing }}$ is  Eq.~\ref{eq:dephasing}, and the $tgate$ denotes the running time of single quantum gate, $T1$ is qubit lifetime, $T2$ is qubit coherence time. 
    \item Depolarizing kraus operator
    \begin{equation}
    \begin{aligned}
    &K_{1}=\sqrt{1-3 p / 4} \times I, K_{2}=\sqrt{p} / 2 \times X \\
    &K_{3}=\sqrt{p} / 2 \times Y, K_{4}=\sqrt{p} / 2 \times Z
    \end{aligned}
    \end{equation}
    \item Bit or phase flip operator
    \begin{equation}
    K_{1}=\left[\begin{array}{cc}
    \sqrt{1-p} & 0 \\
    0 & \sqrt{1-p}
    \end{array}\right], K_{2}=\left[\begin{array}{cc}
    0 & -i \times \sqrt{p} \\
    i \times \sqrt{p} & 0
    \end{array}\right]
    \end{equation}
    \item Phase damping operator
    \begin{equation}
    K_{1}=\left[\begin{array}{cc}
    1 & 0 \\
    0 & \sqrt{1-p}
    \end{array}\right], K_{2}=\left[\begin{array}{cc}
    0 & 0 \\
    0 & \sqrt{p}
    \end{array}\right]
    \end{equation}
\end{itemize}
\begin{lstfloat}[bt]
\begin{lstlisting}[
  style=PythonStyle,
  numbers=left,
  linewidth=0.94\columnwidth,
  xleftmargin=0.06\columnwidth,
  basicstyle=\linespread{1}\scriptsize\ttfamily,
  caption=QPanda noise model example.,
  label=code:QPanda_noise,
]
# Init quantum simulator and allocation quantum and classical register
qvm = NoiseQVM()
qvm.init_qvm()
qubits = qvm.qAlloc_many(4)
cbits = qvm.cAlloc_many(4)

# Set noise to Circuit
# Set bit flip noise to X gate
qvm.set_noise_model(NoiseModel.BITFLIP_KRAUS_OPERATOR, GateType.PAULI_X_GATE, 0.1)
qves = [[qubits[0], qubits[1]], [qubits[1], qubits[2]]]
# set damping noise to CNOT gate
qvm.set_noise_model(NoiseModel.DAMPING_KRAUS_OPERATOR, GateType.CNOT_GATE, 0.1, qves)
# Set readout error 
f0 = 0.9
f1 = 0.85
qvm.set_readout_error([[f0, 1 - f0], [1 - f1, f1]])
# Set rotation error
qvm.set_rotation_error(0.05)
\end{lstlisting}
\end{lstfloat}\par
In code~\ref{code:QPanda_noise}, we can add noise to the circuit with the \colorbox{bkgd}{set\_noise\_model} and \colorbox{bkgd}{NoiseQVM}. Through the interfaces \Colorbox{bkgd}{BITFLIP\_KRAUS\_OPERATOR}, \Colorbox{bkgd}{DAMPING\_KRAUS\_OPERATOR}, etc., to set the noise model.\par 
\subsection{Tools of QPanda}\label{tools}
QPanda aims to serve researchers better, reduce code redundancy, and provide a more convenient research framework, essential tools such as Variational quantum circuits, Pauli operators, Fermions operators, and Hamiltonian quantities, the quantum circuit visualization tools. It is convenient for researchers to observe the circuit. In aid of helping researchers' job execute on OriginQ's QPU, QPanda also provides quantum programs to transform OriginIR; it also supports the transformation of QPanda programs to Qasm and Quil to submit quantum jobs on different QPUs. Moreover, quantum programs implement classical operations, such as while and if operations with Qif and QWhile interfaces. \par
\begin{figure*}[hbt]
\centering
\subfigure[]{
\begin{minipage}[]{0.3\linewidth}
\centering
\includegraphics[height=.06\textheight,width=1\textwidth]{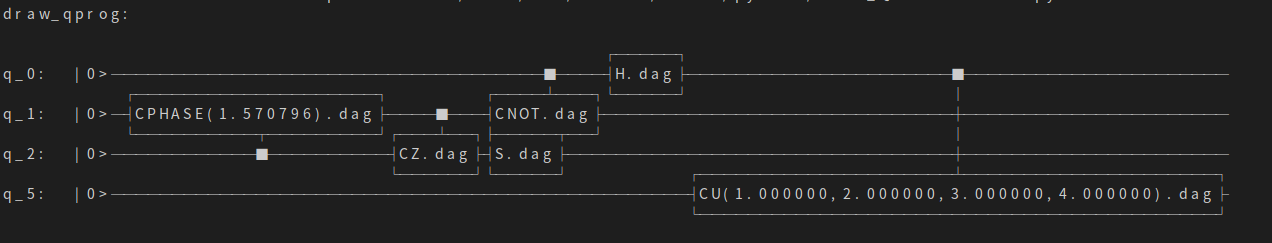}
\end{minipage}
}
\subfigure[]{
\begin{minipage}[]{0.3\linewidth}
\centering
\includegraphics[height=.06\textheight,width=1\textwidth]{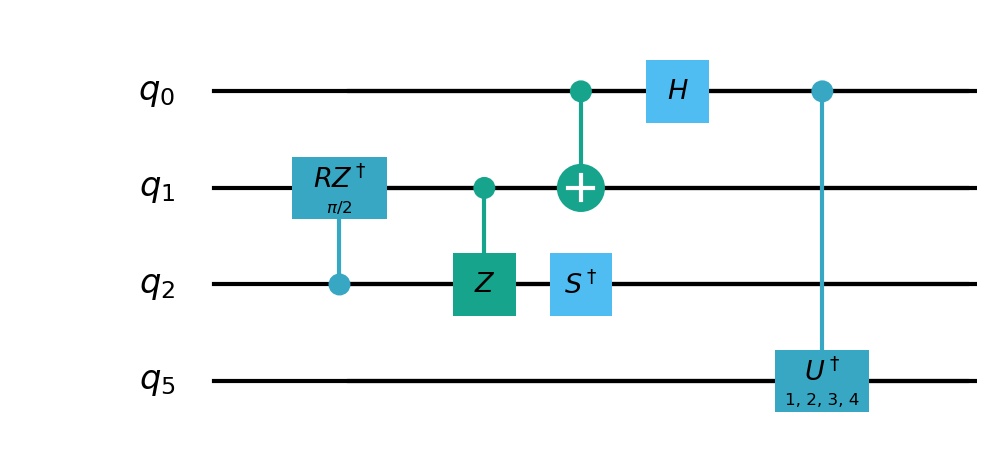}
\end{minipage}
}
\subfigure[]{
\begin{minipage}[]{0.3\linewidth}
\centering
\includegraphics[height=.06\textheight,width=1\textwidth]{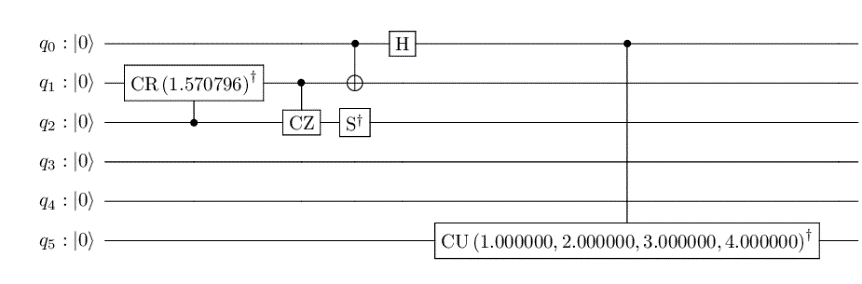}
\end{minipage}
}
\caption{Quantum circuit visualization}
\label{fig:vis}       
\end{figure*}
\begin{lstfloat}[bt]
\begin{lstlisting}[
  style=PythonStyle,
  numbers=left,
  linewidth=0.94\columnwidth,
  xleftmargin=0.06\columnwidth,
  basicstyle=\linespread{1}\scriptsize\ttfamily,
  caption=Variational quantum circuit example.,
  label=code:QPanda_vqc,
]
# Creat variable
x = var(1)  
y = var(2)  
	  
# Generate a variable quantum circuit 
vqc = VQC()  
	  
# Insert variational quantum gate 
vqc <<  (VQG_RX(q[0], x))  
vqc <<  (VQG_RY(q[1], y))  
vqc <<  (VQG_RZ(q[0], 0.12))  

\end{lstlisting}
\end{lstfloat}
\subsubsection{Variational quantum circuit}\label{vqc}
Variational quantum circuits (VQC) are derived from the variational principle in quantum mechanics. The core is optimizing the next iteration by calculating the results obtained from the previous circuit operation to approach the problem's solution continuously. It mainly comprises parameterizable quantum circuits, namely variable subcircuits and measure operations. \par
VQC usually requires external circuits (little gates and qubits) that effectively resist the influence of noise in quantum computers. Therefore, VQC is widespread in NISQ QPU. The variable quantum circuit is a significant feature of VQC. QPanda provides variable quantum gates, parameter variables, and operators to help researchers reconstruct circuits when the parameters are updated. And then rebuild the circuit to optimize parameters by modifying the variable quantum gate parameters, significantly reducing code redundancy. Quantitatively, the algorithm research is accelerated.\par
In Code~\ref{code:QPanda_vqc}, building variable quantum circuits constructs objects through \Colorbox{bkgd}{VQC} interface. Then generates a specific variable class \Colorbox{bkgd}{var} for implementing classical operations (arithmetic operations, matrix operations) and quantum operations (\Colorbox{ bkgd}{qop}, \Colorbox{bkgd}{qop\_measure}, etc. are used to find the expectation of the input Hamiltonian), and store the variables of a specific quantum-classical network. Finally, insert the variable quantum gate (\Colorbox{bkgd}{VQG\_RX}, \Colorbox{bkgd}{VQG\_RY}, etc.) into the variational quantum circuit. And update it continuously to complete the construction of the variational quantum circuit.\par
\subsubsection{Quantum circuit visualization}\label{visualization}
A circuit constructed may require a large number of gates. If the experimental result is not expected, the circuit must be modified. And through the scheme, reviewing the code wastes the researcher's time on some non-critical aspects. Therefore, QPanda has launched a quantum circuit visualization tool to solve this redundancy, which significantly facilitates the researchers' experiments. Then, in Code~\ref{code:QPanda_vis}, the interfaces of printing circuit \Colorbox{bkgd}{print}, and quantum circuit diagram and tex file\Colorbox{bkgd}{draw\_qprog} also be provided. As shown in figure \ref{fig:vis}, (a) is the printed circuit diagram of the console, (b) represents the output quantum circuit diagram in the folder, and (c) denotes the compiled pdf of the generated tex file.\par
\begin{lstfloat}[bt]
\begin{lstlisting}[
  style=PythonStyle,
  numbers=left,
  linewidth=0.94\columnwidth,
  xleftmargin=0.06\columnwidth,
  basicstyle=\linespread{1}\scriptsize\ttfamily,
  caption=Quantum circuit visualization example.,
  label=code:QPanda_vis,
]
# Init quantum simulator and allocation quantum and classical register
machine = CPUQVM()
machine.init_qvm()
qubits = machine.qAlloc_many(2)
prog = QCircuit()      
prog << CU(1, 2, 3, 4, qubits[0], qubits[5]) << H(qubits[0]) << S(qubits[2])\      
<< CNOT(qubits[0], qubits[1]) << CZ(qubits[1], q[2]) << CR(qubits[2], qubits[1], PI/2)      
prog.set_dagger(True)
      
# Console output      
print(prog)    
	        
# Image output    
draw_qprog(prog, 'pic', filename='D:/test_cir_draw.png')  
	  
# Tex output  
draw_qprog(prog,'latex') 
\end{lstlisting}
\end{lstfloat}
\subsubsection{Prog to QASM, Quil}\label{qasm}
QASM and Quil are hardware-oriented assembly languages corresponding to IBM and Quil. QASM is a quantum assembly language proposed by IBM. Quil is a low-level direct description of quantum programs, and its status is similar to the hardware description language or assembly language in classical computers. The interfaces \Colorbox{bkgd}{convert\_qprog\_to\_qasm} and \Colorbox{bkgd}{convert\_qprog\_to\_quil} are used to transform QPanda quantum programs into Qasm and Quil in Code~\ref{code:QPanda_convert}, respectively.\par
\begin{lstfloat}[bt]
\begin{lstlisting}[
  style=PythonStyle,
  numbers=left,
  linewidth=0.94\columnwidth,
  xleftmargin=0.06\columnwidth,
  basicstyle=\linespread{1}\scriptsize\ttfamily,
  caption=QPanda convert prog example.,
  label=code:QPanda_convert,
]
# Init quantum simulator
machine = CPUQVM()
machine.init_qvm()  
	    
# Prog to Qasm 
Qasm = convert_qprog_to_qasm(prog, machine)  
  
# Prog to Quil  
Quil = convert_qprog_to_quil(prog, machine) 

# Prog to OriginIR  
OriginIR = convert_qprog_to_originir(prog, machine)
\end{lstlisting}
\end{lstfloat}
\subsubsection{Qif and QWhile}\label{Qif}
Through the quantum condition operations (\Colorbox{bkgd}{create\_if\_prog},  
\Colorbox{bkgd}{create\_while\_prog}) interfaces to build quantum circuits more flexibly. Explaining classical bits for addition, subtraction, multiplication, division, and XOR operation is necessary to judge boundary conditions.\par
\section{Experiment}\label{experiment}
In this section, we design two experimental directions to fully the simulation performance of QPanda. The first direction is a numerical simulation, which mainly simulates different dimension gates. In addition, we also designed an experiment to verify the optimized performance of the computing backend and circuit and the details in Section~\ref{numerical}. The next direction is the comparative experiment—the details of comparing the performance with the existing quantum lib in Section~\ref{compare_experiment}. \par
\subsection{Numerical performance}\label{numerical}
We complete the numerical simulation experiments in two ways. First, estimate simulating each gate time of QPanda by simulating the different gate dimensions; then, explain that the OpenMP and circuit optimization improve the simulation performance. Here, we show the experimental evidence. The table ~\ref{tab:benchmark} shows the benchmark environment.
\begin{table}[htbp]
    \centering
     \caption{Benchmark environment}
    \begin{tabular}{c|c}
    \hline\hline
    Environment &Version\\
    \hline\hline
     GCC       & 7.3.0\\
     Python    & 3.9.0\\
     System      & CentOS\\
     CPU       & Intel(R) Core(TM) i9-9900K CPU \@ 3.60GHz\\
     GPU    & NVIDIA GEFORCE RTX 3090 \@ 24GB \@ 1.70Ghz \\
     \hline\hline
    \end{tabular}
    \label{tab:benchmark}
\end{table}
\begin{figure}[htbp]
    \centering
    \includegraphics[width=8cm]{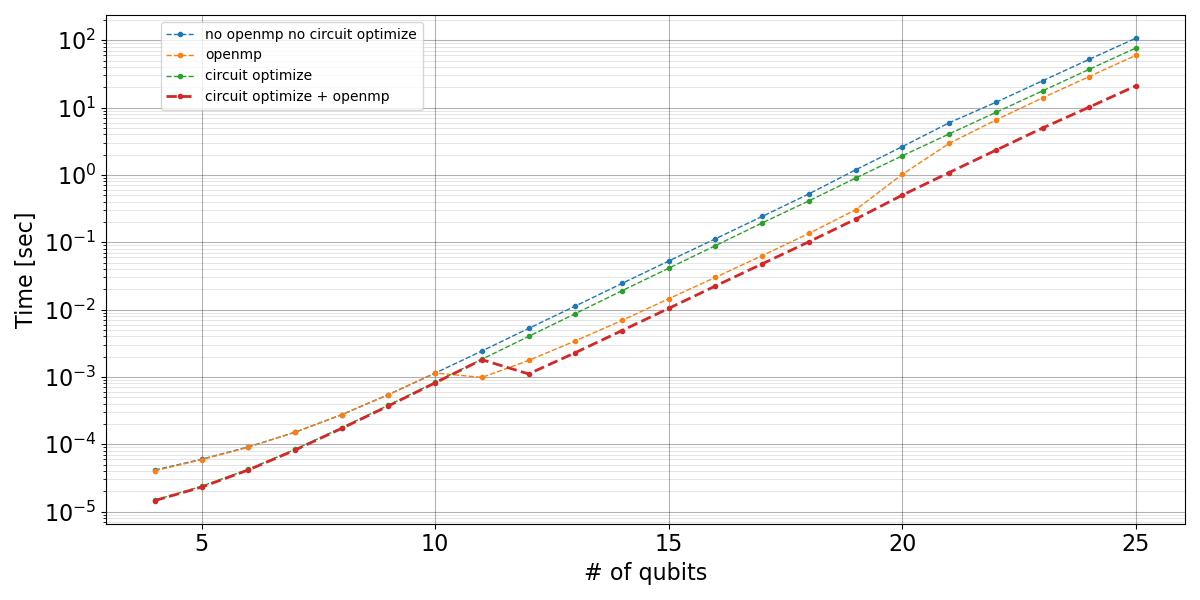}
    \caption{Performance of opening OpenMP, circuit optimization simulates random circuit}
    \label{fig:openmp}
\end{figure}
\begin{figure}
    \centering
    \includegraphics[width=8cm]{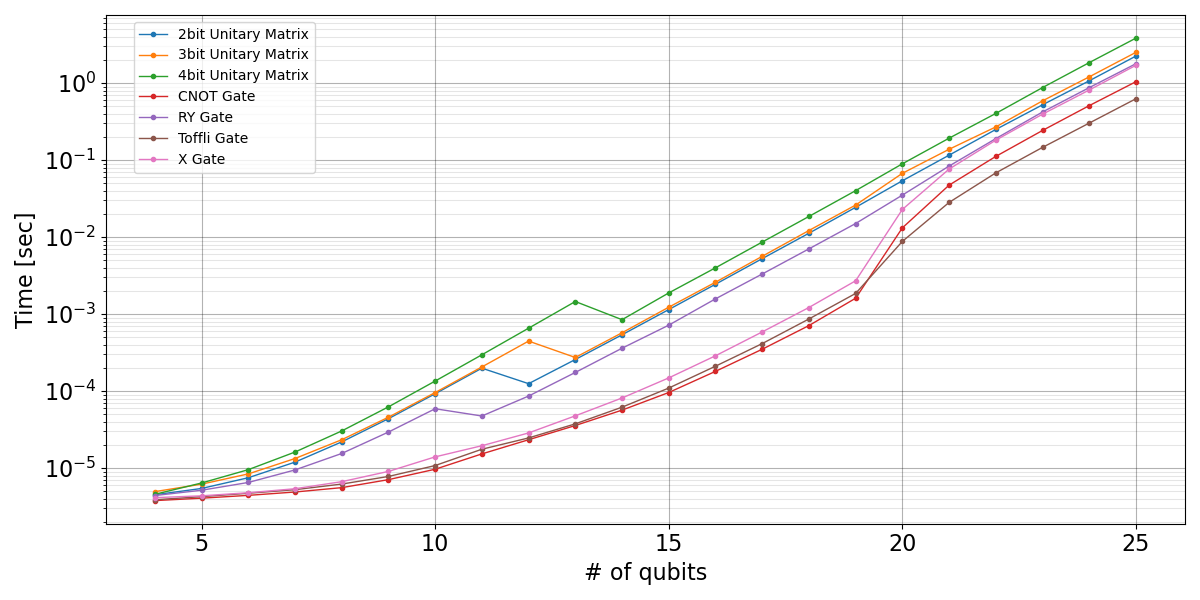}
    \caption{Compare the time of different quantum gates dimensions}
    \label{fig:gate}
\end{figure}
\begin{figure*}
    \centering
    \includegraphics[width=12cm]{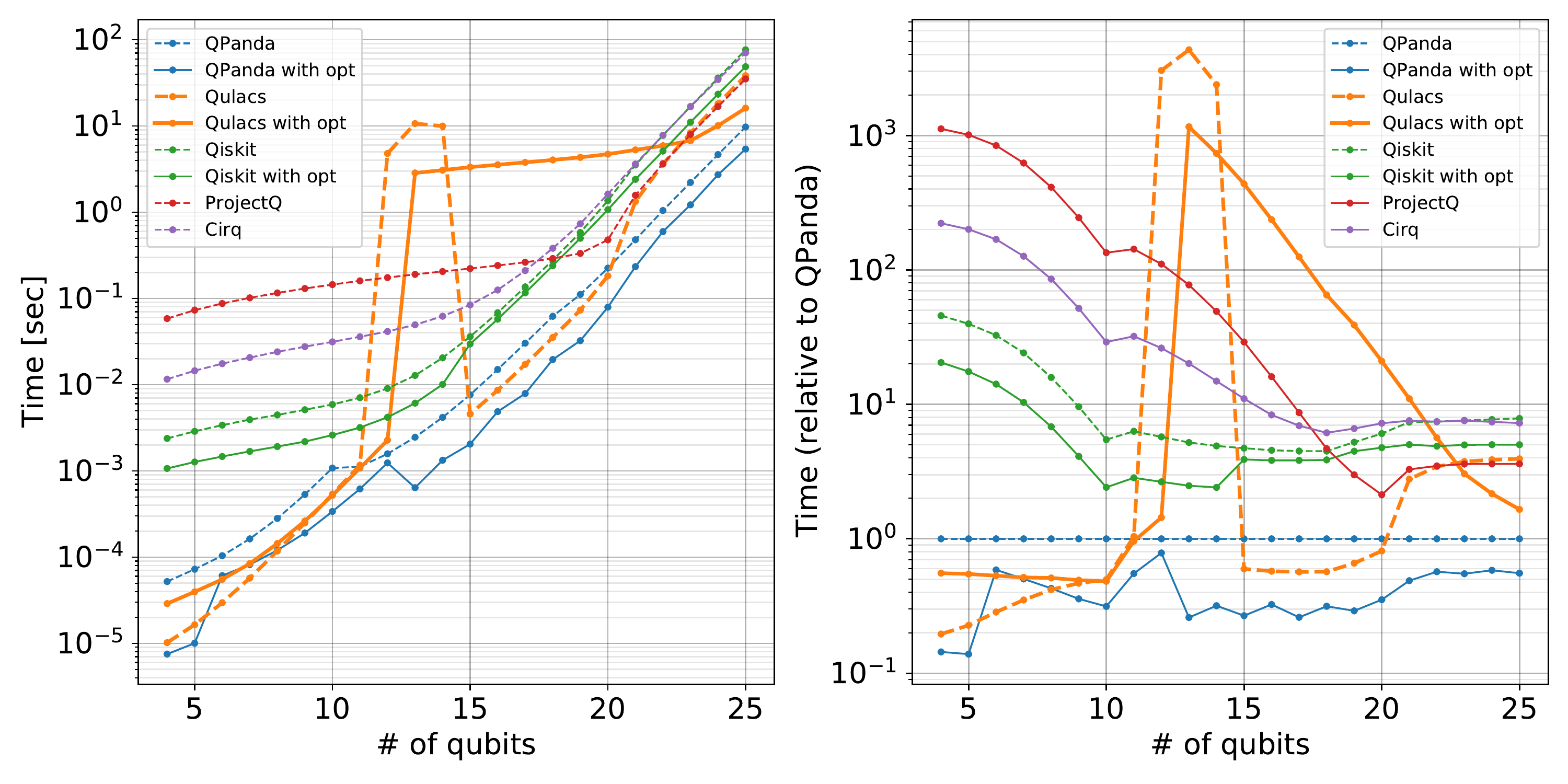}
    \caption{Times for simulating random quantum circuits with a single thread using several libraries}
    \label{fig:single}
\end{figure*}
\begin{figure*}
    \centering
    \includegraphics[width=12cm]{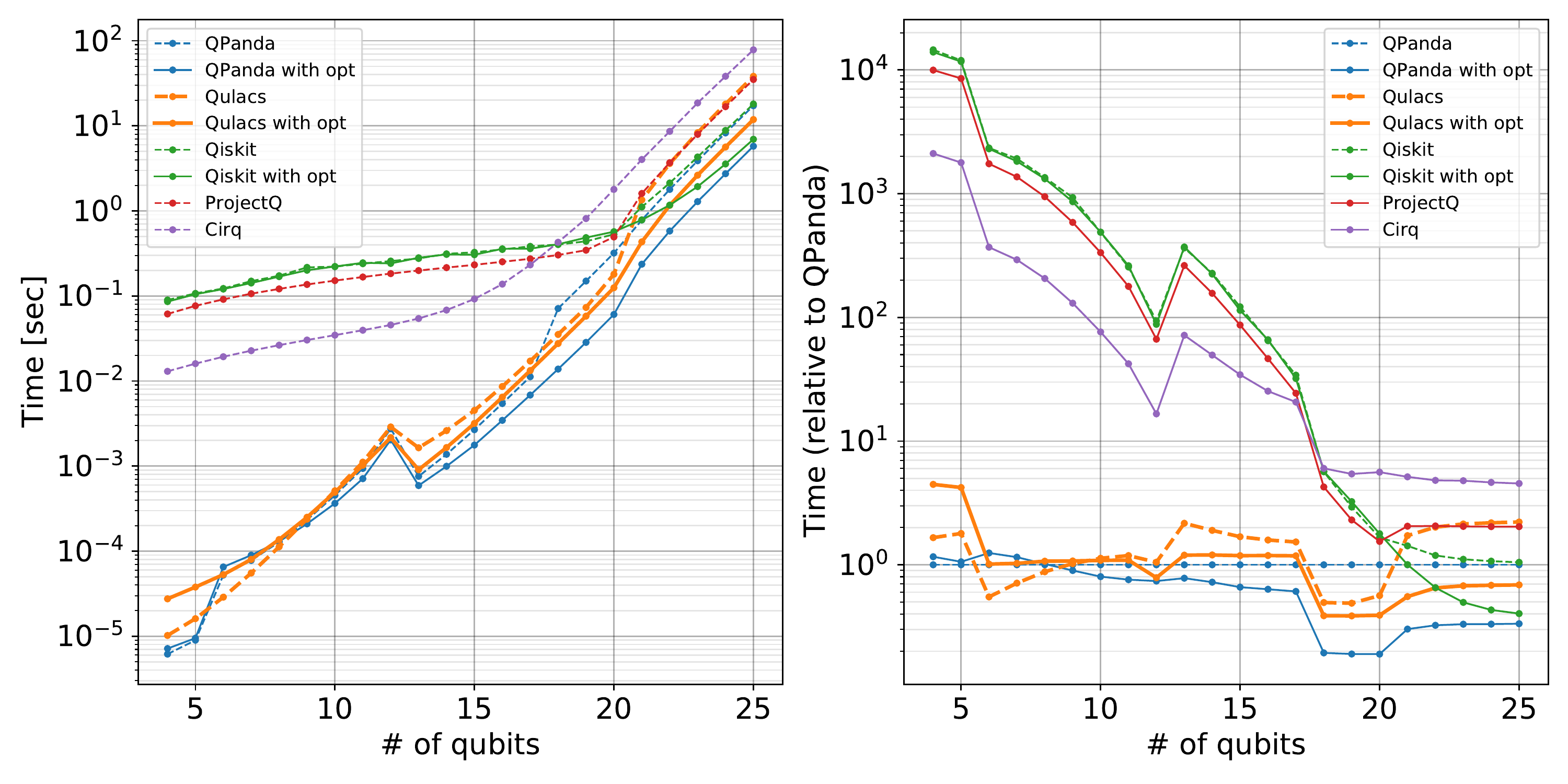}
    \caption{Times for simulating random quantum circuits with Muti-thread using several libraries}
    \label{fig:multithread}
\end{figure*}
\begin{figure*}
    \centering
    \includegraphics[width=12cm]{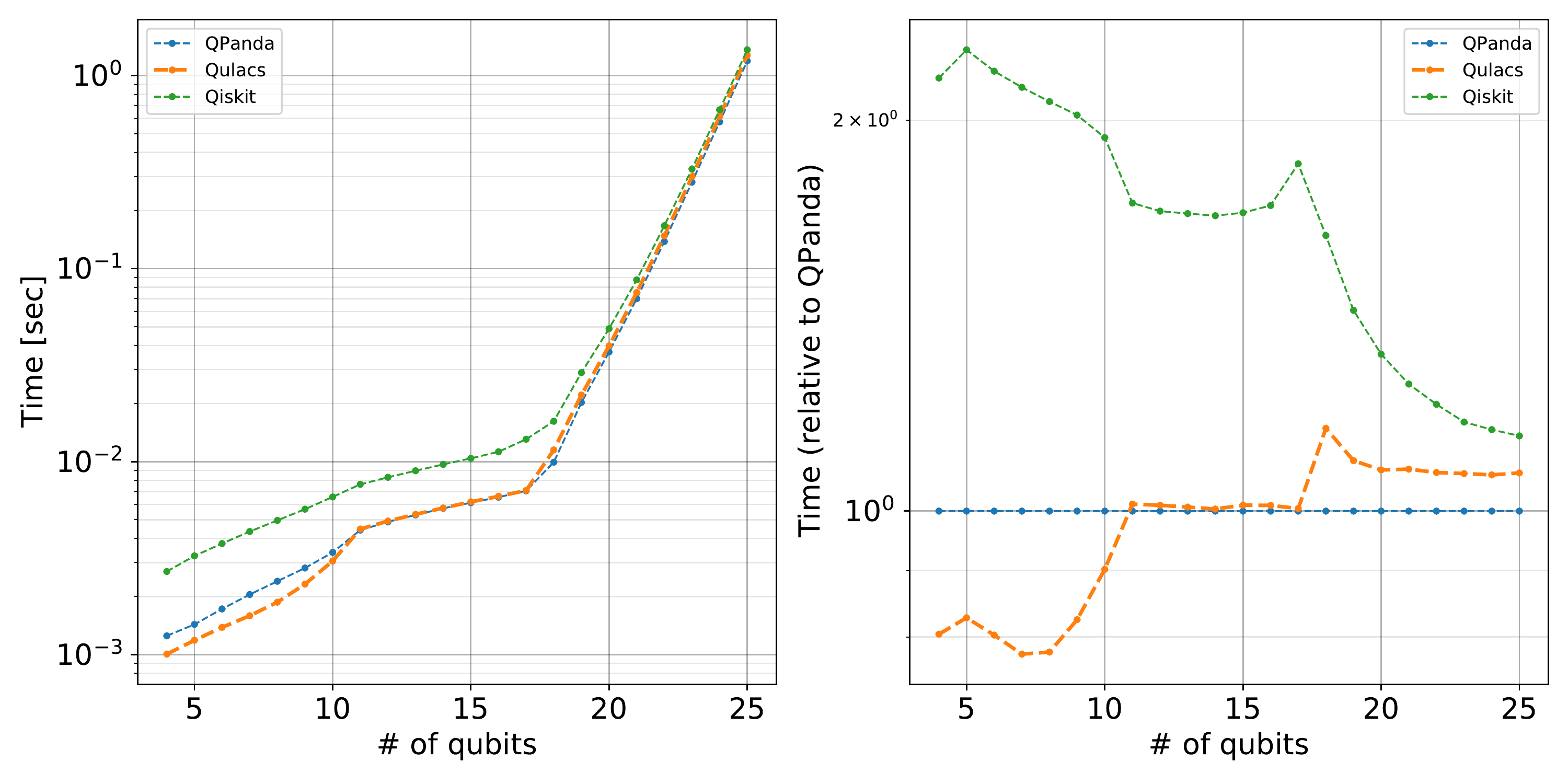}
    \caption{Times for simulating random quantum circuits with GPU using several libraries}
    \label{fig:GPU}
\end{figure*}
\subsubsection{Performance of basic gates}\label{basic_performances}
In this section, we experiment with the simulation performance of different gate dimensions so that researchers estimate the cost of gates accurately. The experiment simulation backend is a full-amplitude simulation backend. Figure\ref{fig:gate} shows the relationship between the gates and qubits; then, all cost time grows exponentially with the qubit increase. Due to the overhead of calling a C++ function from Python, each gate time cost converges to a particular value. When applying little qubits, the gates involved in the experiment are $R_x$, $R_y$, CNOT, Toffoli, and unitary Matrix. Among them, Unitary Matrix dimensions are $2^1\times2^1, 2^2\times2^2, 2^3\times2^3$, and denote 1, 2, and 3 qubits, respectively. It's easy to see that the unitary gate and RY gate will have performance enhancements on 11, 12, 13, and 14 qubits, respectively. Because the adaptive OpenMP of QPanda is turned on, which will improve the simulation performance. Since the Pauli-X, Toffoli, and CNOT gates are optimized during operation, their time cost bottleneck is due to the operational memory rather than arithmetic operation. The Pauli-X, Toffoli, and CNOT gates are the fastest gates. \par

\subsubsection{Performance of OpenMP, circuit optimization}\label{openmp_performances}
To verify the efficient simulation performance of OpenMP and circuit optimization. We compare the time cost between OpenMP, circuit optimization, and no optimization, OpenMP, by simulating a random circuit. The result is shown in figure \ref{fig:openmp}, where the benchmark environment is shown in the table~\ref{tab:benchmark}, using the full amplitude as the simulation backend. From the figure\ref{fig:openmp}, opening OpenMP and circuit optimize options have the best performance of the experiment, and none use, the none use simulation is the worst of all. However, reduce opening unnecessary multithreading so that QPanda's adaptive opening OpenMP, where the performance decreases in 11,12 qubit of OpenMP or circuit optimization.  \par

\subsection{Compared with existing libs}\label{compare_experiment}
In this section, we compare the simulation performance between CPU and GPU, and the benchmark is selected as a random circuit, the structure of it as two layers. The detail is as follow:\par
1. Rotation angle layer: consists of $R_X$, $R_Y$, $R_Z$ gate.\par
2. CNOT layer: The target qubit of the CNOT gate is the $i$th qubit, and the control qubit is the $((i+1)\ mod\ n)$ qubit, where $(0 \leq i \leq n)$.\par
Here, all initial quantum states are $|0\rangle$ states on Pauli-Z basis, and the other details of this random circuit are as follows in \cite{arute2019quantum}. This experiment is via benchmark of pytest\cite{krekel2004pytest}. And the lib and hardware version as shown in table~\ref{tab:benchmark},~\ref{tab:version}. The libs' version compared in this paper is the latest October 2022. Since the whole simulation includes the building, compilating, and executing, all involved in the compared assessment, the circuit's entire execution time is regarded as the evaluation index. If the lib provides a circuit optimization option, also compared in this benchmark. The note is that we add the optimized time to the total time for comparison.\par
\begin{table}[htbp]
    \centering
    \caption{A list of libraries and versions for CPU benchmark}
    \begin{tabular}{c|c}
      \hline\hline
      Library   & Version \\
      \hline\hline
      GCC       & 7.3.0\\
      Python    & 3.9.0\\
      NumPy       & 1.23.3\\

      MKL       & 2.4.0\\
      \hline\hline
      ProjectQ       & 0.7.3\\

      Qiskit       & 0.38.0\\

      Qiskit Aer       & 0.11.0\\

      Qiskit Terra       & 0.21.2\\

      Qulacs       & 0.5.1\\

      Pyqpanda       & 3.7.14\\

      Cirq      & 1.0.0\\
      \hline\hline
    \end{tabular}
    \label{tab:version}
\end{table}

To ensure the objectiveness and fairness of the experiment, this paper selects the simulation backend that full-amplitude method of all compare libs. Among them, We use the \Colorbox{ bkgd}{StatevectorSimulator} backend to simulate because it's considered to be the fastest simulation backend of Qiskit\cite{aleksandrowicz2019qiskit}. Since Qiskit\cite{aleksandrowicz2019qiskit} also provides circuit optimization option\Colorbox{ bkgd}{fusion}, we will compare it in the experiment. Due to the benchmark circuit without measurement operation, ProjectQ\cite{steiger2018projectq} will throw a warning, which may lead to an extra benchmark cost. However, this extra cost is constant so that it won't influence the performance comparison and via \Colorbox{bkgd}{MainEngine} to simulate. In Cirq\cite{quantum2020team}, this paper uses \Colorbox{bkgd}{simulator.simulate} to simulate the benchmark circuit. Then Qulacs\cite{suzuki2021qulacs} uses its default simulator backend\Colorbox{bkgd}{update\_quantum\_state}, and this lib also provides circuit optimization option\Colorbox{bkgd}{QCO}, which will be included in the experiment. Finally, QPanda via \Colorbox{bkgd}{CPUQVM} to complete the benchmark and support the circuit optimization option\Colorbox{bkgd}{aggregate\_operations}.\par
\begin{table*}
    \centering
    \caption{QAOA program compile and optimize in different toplogies}
    \begin{tabular}{c|c|c|c|c|c}
    \toprule[1.5pt]
    \multirow{2}{*}{Algorithm}    &\multirow{2}{*}{Topologies} & \multicolumn{2}{|c|}{Compiled circuit} & \multicolumn{2}{|c}{optimized circuit}\\
    \cline{3-6}
    & &Circuit depth &CNOT numbers &circuit depth &CNOT numbers
    \\ \hline\hline
    \multirow{2}{*}{4-qubits-QAOA} &Grid  &82 &122&\textbf{80}&\textbf{86}\\
                                   &Circle &130&269&\textbf{55}&\textbf{62}\\
    \hline
    \multirow{2}{*}{6-qubits-QAOA} &Grid  &189 &327&\textbf{143}&\textbf{204}\\
                                   &Circle &304&627&\textbf{173}&\textbf{204}\\
    \hline
    \multirow{2}{*}{8-qubits-QAOA} &Grid  &296 &650&\textbf{255}&\textbf{470}\\
                                   &Circle &485&1145&\textbf{358}&\textbf{656}\\
    \hline
    \multirow{2}{*}{10-qubits-QAOA} &Grid  &480 &1025&\textbf{428}&\textbf{872}\\
                                   &Circle &758&1994&\textbf{635}&\textbf{1335}\\
    \hline
    \multirow{2}{*}{12-qubits-QAOA} &Grid  &659 &1566&\textbf{584}&\textbf{1200}\\
                                   &Circle &1048&2805&\textbf{831}&\textbf{1818}\\
                                   \hline
    \multirow{2}{*}{14-qubits-QAOA} &Grid  &1174 &2474&\textbf{792}&\textbf{1736}\\
                                   &Circle &1571&4127&\textbf{1159}&\textbf{2717}\\
                                   \hline
    \multirow{2}{*}{16-qubits-QAOA} &Grid  &1355 &3011&\textbf{1011}&\textbf{2225}\\
                                   &Circle &2669&5744&\textbf{1977}&\textbf{3938}\\
    \hline\hline
    \end{tabular}
    \label{Tab:QPU}
\end{table*}
Firstly, we show the benchmark for CPU, in which the experiment is shown as single-thread and multi-thread, respectively. The single-thread and multi-thread benchmarks are shown in \ref{fig:single} and \ref{fig:multithread}, respectively. The dotted and solid lines are used to indicate whether the QPanda, Qulacs\cite{suzuki2021qulacs}, Qiskit\cite{aleksandrowicz2019qiskit} libs are opening optimized options or not. Due to the qubits increase, the arithmetic costs increase exponentially, and whole lib's benchmarks have similar performance after 20 qubits without optimizing. However, QPanda has the best performance when opening the optimizing option because QPanda adopts various circuit optimization schemes. The gates will fuse rapidly, decreasing arithmetic costs.\par
Here, we select three libs to compare the performance with GPU optimization, QPanda,  Qulacs\cite{suzuki2021qulacs}, Qiskit\cite{aleksandrowicz2019qiskit}. Then, to better test the pure optimize effect, we compare other libs without circuit optimization. From the GPU result in figure\ref{fig:GPU}, The simulation performance of GPU is much better than CPU. The optimization effect of whole libs is almost similar, where the gap is only in the accuracy is $10^{-3}$.\par
Finally, Through the numerical simulation demonstrated the optimized schemes are more efficient. And then, the CPU and GPU benchmarks prove the high performance of QPanda simulation, and the performance is top of compared libs.\par
\subsection{Physical qubit mapping and optimizing}
In this section, we will introduce the circuit mapping and optimizing method of the requirement on QPU. Therefore, from the table ~\ref{Tab:QPU} we show the result of circuit mapping on the cycle or grid topology and the optimized result. We can see that the optimized result's circuit depth and CNOT numbers are shorter than the pre-optimizing result. \par
Unfortunately, quantum computing is subject to the influence of qubit decoherence time and chip topology in the current NISQ era, so we cannot directly execute the original quantum program on QPU. Instead, we need to use a physical bit mapping scheme. Then, to improve the circuit's fidelity, we also need to reduce the redundant gates in the circuit. However, it's a powerful embodiment of the fidelity of quantum computers. So, The mapping and optimizing interface provided by QPanda is certainly exciting.\par

\section{Conclusion and feature work}\label{conclusion}
This paper introduces the quantum programming framework-QPanda. It proposes a design concept for applying multi-scenarios and supports the research of quantum finance, quantum chemistry, quantum fluid mechanics, etc. At the same time, to solve the hybrid computing problem, the fusion scheme of quantum computers and supercomputers is adopted, which is applied in quantum machine learning and quantum error mitigation. Finally, the noise simulation backend is introduced for the robustness research of quantum algorithms.\par
\begin{itemize}
    \item In the benchmark test of random circuits, QPanda is superior to other state-of-the-art frameworks in terms of single-thread CPU and multi-thread CPU. Moreover, GPU optimization also has a better performance.
    \item Through the QPU experiment, it is found that the compiled and optimized quantum circuit will have a high fidelity when executed in QPU.
\end{itemize}
 In the future, to be more convenient for researchers to use QPanda, we will also provide QPanda as a backend for other libs, such as Cirq\cite{quantum2020team}, PennyLane\cite{bergholm2018pennylane}.\par
 
\bibliographystyle{ieeetr}
\bibliography{QPanda.bib}

\end{document}